\documentclass[journal,a4paper,10pt]{IEEEtran}

\usepackage[latin1]{inputenc}
\usepackage{times,amsmath,amsfonts}
\usepackage{pstool}
\usepackage{subfigure}
\usepackage{multirow}
\usepackage{enumerate}
\usepackage{graphicx}
\usepackage{MnSymbol}
\usepackage{stfloats} 
\usepackage[table]{xcolor}
\usepackage[square, comma, sort&compress, numbers]{natbib}
\usepackage{algorithm,algorithmic}

\graphicspath{ {Figures/} }

\begin{document}

\title{Channel Impulse Response-based Physical Layer Authentication in a Diffusion-based Molecular Communication System}

\author{
\IEEEauthorblockN{Sidra\ Zafar*, Waqas\ Aman*, Muhammad\ Mahboob\ Ur\ Rahman, Akram\ Alomainy, and Qammer\ H.\ Abbasi}
}

\maketitle

\long\def\symbolfootnote[#1]#2{\begingroup%
\def\thefootnote{\fnsymbol{footnote}}\footnote[#1]{#2}\endgroup}
\symbolfootnote[0]{\hrulefill \\
Sidra Zafar is with the Computer science department, Lahore College for Women University, Lahore, Pakistan (sidzafar.88@gmail.com). \\
Waqas Aman and Muhammad Mahboob Ur Rahman are with the Electrical engineering Department, Information Technology University, Lahore 54000, Pakistan (\{waqas.aman, mahboob.rahman\}@itu.edu.pk). \\
Akram Alomainy is with the School of Electronic engineering and Computer science, Queen Mary University of London, UK (a.alomainy@qmul.ac.uk).\\
Qammer H. Abbasi is with the Department of Electronics and Nano engineering, University of Glasgow, Glasgow, UK (qammer.abbasi@glasgow.ac.uk). \\
* implies equal contribution by the authors.
}

\maketitle

\begin{abstract} 

Consider impersonation attack by an active malicious nano node (Eve) on a diffusion based molecular communication (DbMC) system---Eve transmits during the idle slots to deceive the nano receiver (Bob) that she is indeed the legitimate nano transmitter (Alice). To this end, this work exploits the 3-dimensional (3D) channel impulse response (CIR) with $L$ taps as device fingerprint for authentication of the nano transmitter during each slot. Specifically, Bob utilizes the Alice's CIR as ground truth to construct a binary hypothesis test to systematically accept/reject the data received in each slot. Simulation results highlight the great challenge posed by impersonation attack--i.e., it is not possible to simultaneously minimize the two error probabilities. In other words, one needs to tolerate on one error type in order to minimize the other error type. 

\end{abstract}

\section{Introduction}
\label{sec:intro}

In a diffusion based molecular communication (DbMC) system, communication (transport of molecules from nano transmitter to nano receiver) takes place through a diffusion paradigm. Thus, the DbMC channel is a broadcast channel which implies that the DbMC channel is prone to various kinds of active and passive attacks by the adversaries in the nearby vicinity \cite{Dressler:NanoCommNet:2012}. Though there have been proposals to secure the DbMC systems via cryptographic protocols \cite{Dressler:NanoCommNet:2012}, we are of the opinion that physical layer security, being light-weighted, also makes a strong case for DbMC systems. In this work, inspired by the channel impulse response (CIR) based authentication \cite{Ammar:VTC:2017S} in traditional wireless networks, we propose to exploit the 3-dimensional (3D) CIR as device fingerprint for authentication of nano transmitter in a DbMC system. In a related earlier work, the co-authors studied the authentication problem for an on-body, nano network operating in terahertz band \cite{Mahboob:Access:2017}.

\section{System Model \& Channel Model}
\label{sec:sm}

\subsection{System Model}
\begin{figure}[ht]
\begin{center}
	\includegraphics[width=3in]{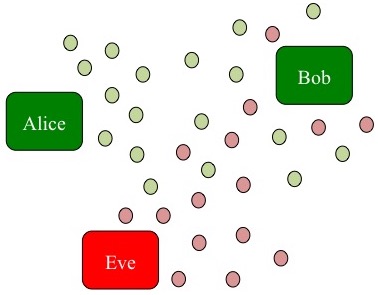} 
\caption{System model: Eve emits same kind of molecules as Alice, but Eve transmits during time-slots not utilized by Alice in order to stay undetected. The molecules emitted by Alice (Eve) are shown in green (red). }
\label{fig:sysmodel}
\end{center}
\end{figure}

We consider a scenario whereby a legitimate nano transmitter (Alice) talks to a legitimate nano receiver (Bob), while an active malicious node (Eve) is present nearby (see Fig. \ref{fig:sysmodel}). We assume that the DbMC channel is time-slotted. We further assume that nano transmitters use pulse-based modulation (i.e., on/off keying) whereby Alice (Eve) releases $Q_A$ ($Q_E$) molecules during each slot. We consider impersonation attack whereby Eve pretends to be Alice in order to inject malicious data into Bob's system while staying undetected. In other words, Eve is a clever impersonator which does spectrum sensing in order to transmit in those slots when Alice is idle. Finally, to deceive Bob, Eve releases the same kind of molecules as Alice.

\subsection{The DbMC Channel Model}

\begin{figure}[ht]
\begin{center}
	\includegraphics[width=3.8in]{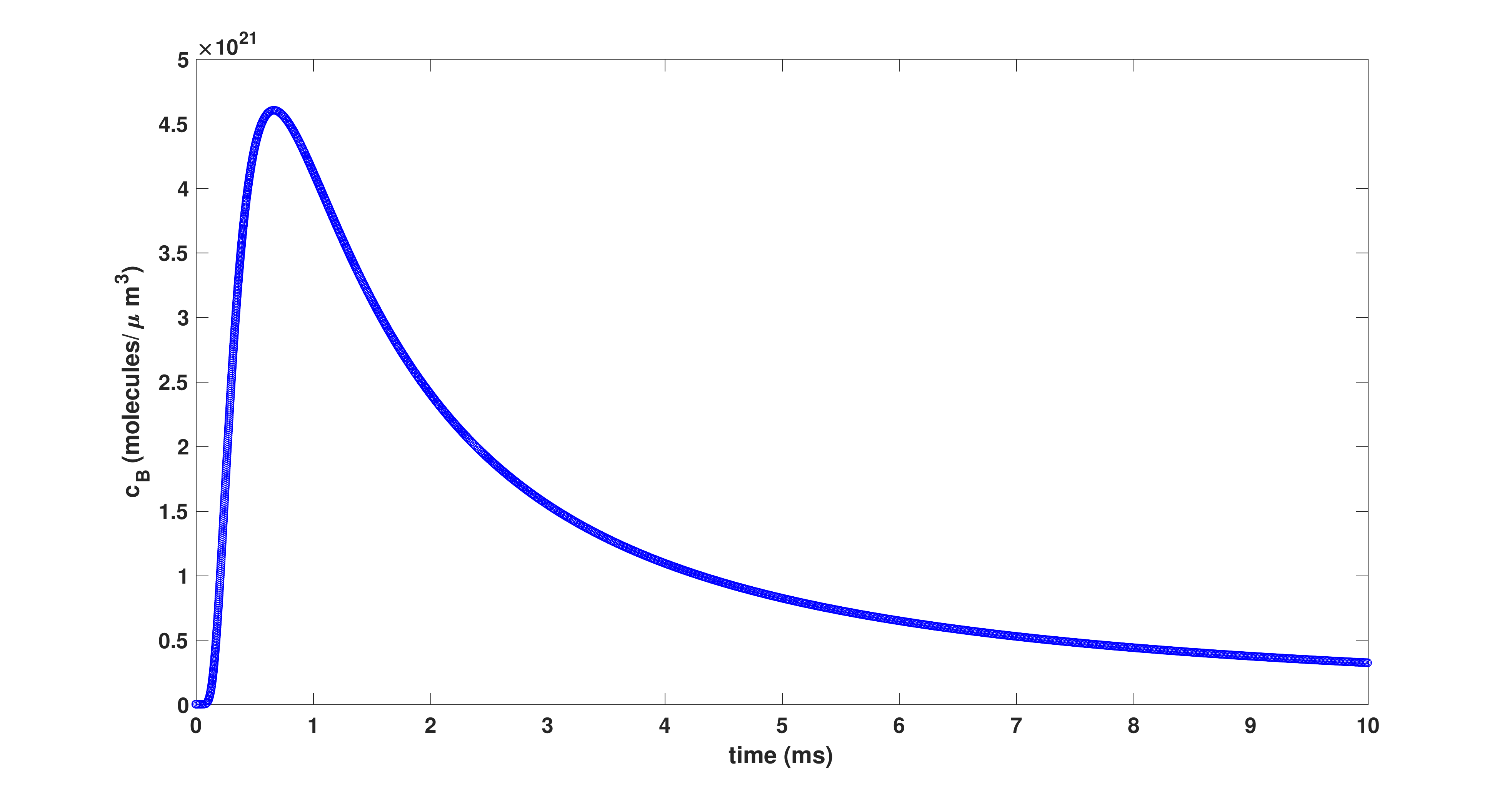} 
\caption{The received molecular pulse at Bob (for $Q_A=5\times 10^5$).}
\label{fig:channel-model}
\end{center}
\end{figure}

Let $c_B(d,t)$ represent the molecule concentration at Bob (which is at distance $d_{AB}$ from Alice at time $t$):  
\begin{equation} \label{eq:Fick_sol}
	c_B(d_{AB},t) = \frac{Q_A}{(4\pi Dt)^{\frac{3}{2}}} e^{-\frac{d_{AB}^2}{4Dt}} 
\end{equation}
where $D$ is the diffusion coefficient of the fluid medium. $c_B(d,t)$ is also known as the 3D CIR of the DbMC channel.
A typical DbMC CIR as seen by Bob is shown in Fig. \ref{fig:channel-model}.

\section{The Proposed Method}
\label{sec:method}

During $k$-th timeslot, either Alice, or, Eve will utilize the DbMC channel (assuming that the Eve avoids collisions so as to stay undetected). We assume that Bob utilizes the initial fraction of each time-slot to measure the CIR of the channel occupant. Thus, Bob obtains a noisy measurement $\mathbf{z}(k)$ $\in \mathbb{R}^{L \times 1}$ of the CIR $\mathbf{h}$ $\in \mathbb{R}^{L \times 1}$ during slot $k$ as follows:
\begin{equation}
\label{eq:z_n}
\mathbf{z}(k) = \mathbf{h} + \mathbf{v}(k)
\end{equation}
where $L=L_A=L_B$ is the number of (symbol-spaced) taps of the DbMC CIR and $\mathbf{v}(k)$ $\in \mathbb{R}^{L \times 1}$ is the measurement noise. Specifically, $\mathbf{v}(k) \sim \mathcal{N}(\mathbf{0},\mathbf{\Sigma})$ where $\mathbf{\Sigma}=\sigma^2 (\mathbf{S}^H\mathbf{S})^{-1}$ $\in \mathbb{R}^{L \times L}$; $\sigma^2$ denotes the variance/power of the additive Gaussian noise at Bob\footnote{$\mathbf{S}$ is the matrix formed by the training symbols. See \cite{Ozair:TNB:2019} for more details.}. 

With $\mathbf{{z}}$ available, Bob casts the authentication problem as the following binary hypothesis testing problem:

\begin{equation}
	\label{eq:H0H1_raw}
\begin{cases} 
H_0: & 
\mathbf{{z}}=\mathbf{h_{AB}}+\mathbf{{v}}
\\
H_1: & 
\mathbf{{z}}=\mathbf{h_{EB}}+\mathbf{{v}}
\end{cases}
\end{equation}
Then, ${\mathbf{z}|H_0} \sim \mathcal{N}({\mathbf{h_{AB}}}, \mathbf{\Sigma})$ and ${\mathbf{z}|H_1} \sim \mathcal{N}({\mathbf{h_{EB}}}, \mathbf{\Sigma})$. If $H_0$ ($H_1$) is true, then the data received on the DbMC channel is accepted (rejected) by Bob.

Next, assuming that Bob knows $\mathbf{h_{AB}}$ (and that the DbMC CIR is time-invariant), it applies the following test:
\begin{equation}
	\label{eq:test_raw}
	 (\textbf{z} - \mathbf{h_{AB}})^H \mathbf{\Sigma}^{-1} (\textbf{z} - \mathbf{h_{AB}}) \underset{H_0}{\overset{H_1}{\gtrless}} \delta  
\end{equation}
where $\delta$ is the comparison threshold, a design parameter whose value is to be determined. This work follows Neyman-Pearson procedure to systematically compute the threshold $\delta$. Specifically, $\delta$ is computed by pre-specifying maximum Type-1 error (i.e., probability of false alarm, $P_{fa}$) that Bob can tolerate. Then, for a given Type-1 error, Neyman-Pearson method guarantees to minimize the Type-2 error (i.e., probability of missed detection, $P_{md}$).   

Let $T$ represent the test statistic in Eq. (\ref{eq:test_raw}); i.e., $T=(\textbf{z} - \mathbf{h_{AB}})^H \mathbf{\Sigma}^{-1} (\textbf{z} - \mathbf{h_{AB}})$. Then, $T|H_0 \sim \chi^2(2L)$; i.e., $T$ has central Chi-squared distribution with $2L$ degrees of freedom, under $H_0$. Then, the probability of false alarm $P_{fa}$ (i.e., incorrectly identifying Alice's data as if it is from Eve) is:
\begin{equation} \label{eq:pfa_fc}
	P_{fa} = Pr(T>\delta |H_0) = \int_{\delta}^\infty p_{T|H_0}(x) dx
\end{equation}
where $p_{T|H_0}(x)$ is the probability density function of $T|H_0$. Thus, one can set $P_{fa}$ to a desired value $\alpha$ in Eq. (\ref{eq:pfa_fc}) and solve for the threshold $\delta$.

With $P_{fa}=\alpha$, the performance of the hypothesis test in Eq. (\ref{eq:test_raw}) is solely characterized by the probability of missed detection (success probability of Eve): $P_{md}=Pr(T<\delta |H_1)$. Since computing the distribution of $T|H_1$ is quite involved, we numerically compute the value of $P_{md}$ in simulation section.

\section{Numerical Results}
\label{sec:results}

\begin{figure}[ht]
\begin{center}
	\includegraphics[width=3.8in]{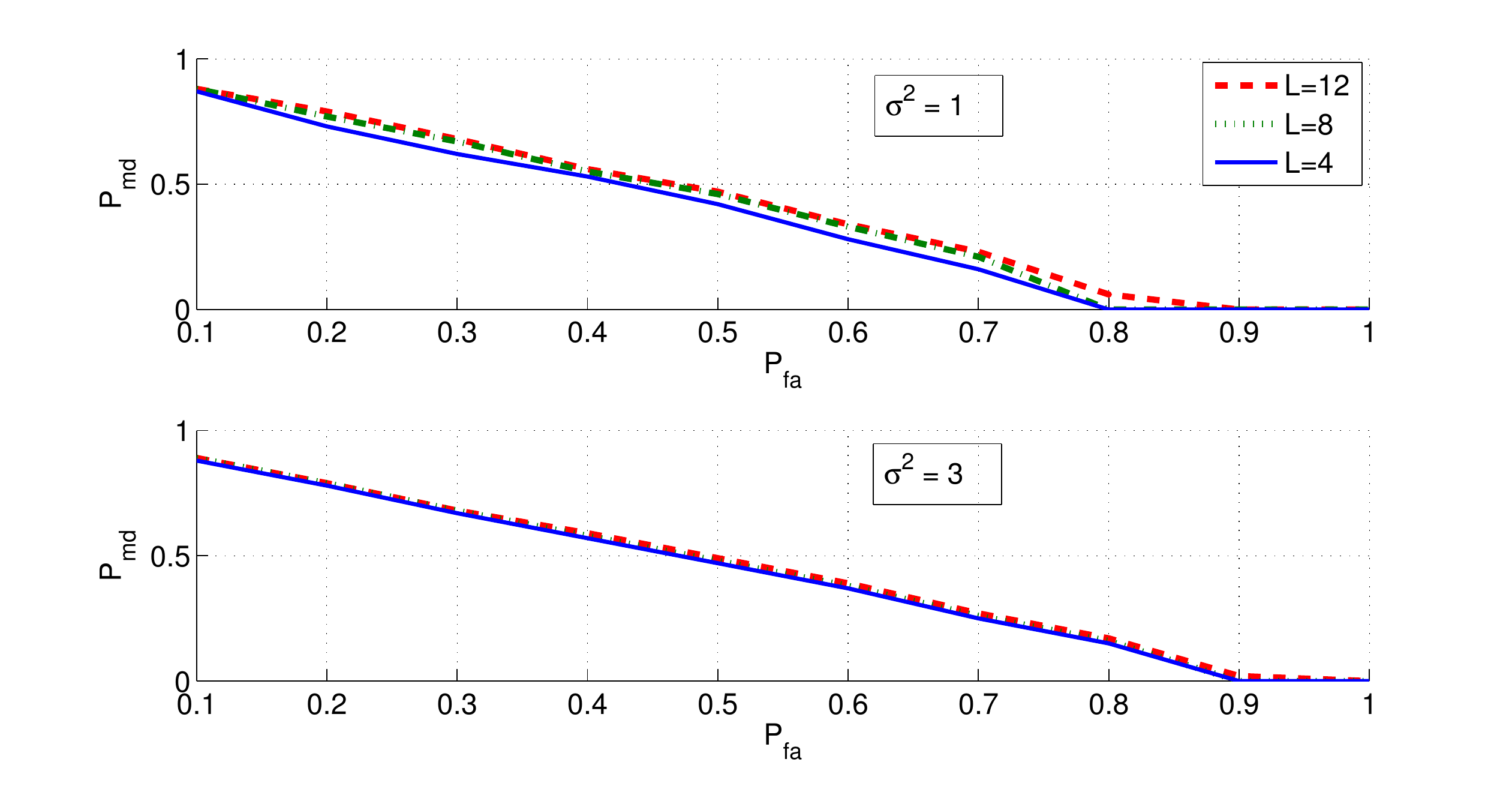} 
\caption{The ROC curves for $\sigma^2=1$ and $\sigma^2=3$.}
\label{fig:roc}
\end{center}
\end{figure}

We set $Q_A=Q_E=5\times 10^5$. We consider DbMC CIR with $L=4,8,12$ taps. larger values of $L$ imply a DbMC channel with shorter time-slots, which in turn implies a DbMC channel with more-pronounced inter symbol interference (ISI) effect. Fig. \ref{fig:roc} plots the receiver operating characteristic (ROC) curves for two different values of variance $\sigma^2$ of the additive noise at Bob. Fig. \ref{fig:roc} signifies that it is not possible to simultaneously minimize the two error probabilities $P_{fa}$ and $P_{md}$. In other words, one needs to tolerate on one error type in order to minimize the other error type. 

\section{Conclusion}
\label{sec:conclusion}

This preliminary work exploited the DbMC CIR as the device fingerprint for authentication of the of nano transmitter during each slot. The proposed method is well-suited to DbMC systems because it is light-weighted (compared to cryptographic techniques which are computationally intensive).

\appendices


\footnotesize{
\bibliographystyle{IEEEtran}

}

\vfill\break

\end{document}